# Boundary conditions effects on the ground state of a two-electron atom in a vacuum cavity

A. Tolokonnikov

Department of Quantum Theory and High Energy Physics, Faculty of Physics,
M.V.Lomonosov Moscow State University, Moscow 119991, Russia
e-mail: tolokonnikov@physics.msu.ru
(Dated: January 22, 2014)

**Abstract:** The ground state properties of the two-electron atom with atomic number $Z \geq 2$ in the spherical vacuum cavity with general boundary conditions of "not going out" are studied. It is shown that for certain parameters of the cavity such atom could either decay into the one-electron atom with the same atomic number and an electron or be in stable state with the binding and ionization energies several times bigger than the same energies of the free atom. By analogy with the Wigner-Seitz model of metallic bonding, the possibility of the existence of such effects on the lattice formed by the vacuum cavities filled with the two-electron atoms of the same type is discussed.

## 1. Introduction

The study of two-electron atom is of particular interest because it is the simplest example of the quantum system with the Coulomb repulsion between electrons. For atomic number $Z \geq 2$ the ground state energy of the free two-electron atom was already obtained with good accuracy in 1928 by Hartree [1] and over the next few years the further researches made by Bethe [2] and Hylleraas [3] allowed to obtain the ground state energy for $Z = 1$. The development of new technologies and experimental techniques has triggered intensive theoretical studies of the two-electron atom in the vacuum microcavity [4-14].

In this work the ground state properties of the two-electron atom with $Z \geq 2$ in a vacuum microcavity with the most general conditions of "not going out" through the cavity boundary are considered. More specific, the influence of such boundary conditions on the behavior of the ground state energy as a function of cavity spatial parameters is studied. Now it should be noted that the term "not going out" emphasizes that such boundary conditions are not necessarily the result of actual confinement of electrons inside the microcavity. Indeed, the general boundary conditions of "not going out" do not require the obligatory vanishing of a wavefunction at the cavity boundary unlike the case of atoms trapped in the cavity by an impermeable or partially permeable potential barrier imitating the compression mode by the external pressure [4-5, 7-14]. Such a situation is realized in the range of actual problems of quantum chemistry and condensed matter physics [4, 15-24]. In particular, in some cases the Neumann conditions imposed on the wavefunction on the cavity boundary could provide not only the state when electrons are localized inside the cavity but also the periodic continuation of the electronic wavefunction as in the Wigner-Seitz model of alcaline metal [21]. Thus, in ordered medium with long-range order, where microcavities could form a spatial lattice, each such cavity with the Neumann boundary conditions corresponds to the Wigner-Seitz cell. When the whole lattice of cells is occupied by atoms of the same type, atomic electrons find themselves in a periodic potential and so the description of their ground state could be based on the principles of the Wigner-Seitz model. Dependence of the ground state properties of the one-electron atom in a cavity on the general boundary conditions of "not going out" was investigated in [22-24]. The rest of this work is organized as follows: in Section 3 the two-electron atom is considered in the spherical cavity with $\delta$-like potential at the boundary which corresponds to the third type boundary conditions imposed on the wavefunction on the cavity boundary and in Section 4 more realistic case, where the "not going out" state of atomic electrons is provided by means of an outer potential shell of nonvanishing width with the Neumann conditions imposed on the wavefunction on its outer boundary, is described. In particular, it will be shown for both cases that in the cavity with the size of the order of the half Bohr radius the two-electron atom could either be in the ground state with the binding and ionization energies several times bigger than the same energies of the free atom or decay into the one-electron atom and a free electron. For the more realistic case the situation where such effects could be observed on the lattice formed by the cavities with the Neumann boundary conditions is discussed.

## 2. Confinement of the Two-particle Quantum System in a Cavity

The nonrelativistic energy functional of two quantum particles with mass $m$, which are confined in a cavity $\Omega$ with boundary $\Sigma$ by means of $\delta$-like potential, could be written as follows

$$E[\Psi] = \int_\Omega dV_1 dV_2 \left( \frac{\hbar^2}{2m} |\vec{\nabla}_1 \Psi|^2 + \frac{\hbar^2}{2m} |\vec{\nabla}_2 \Psi|^2 + \{ \tilde{U}(\vec{r}_1) + \tilde{U}(\vec{r}_2) + W(\vec{r}_1, \vec{r}_2) \} |\Psi|^2 \right) \quad (1)$$

where

$$\tilde{U}(\vec{r}) = U(\vec{r}) + \frac{\hbar^2}{2m} \lambda(\vec{r}) \delta_\Sigma(\vec{r}) \quad (2)$$

Here $U(\vec{r})$ is the potential field inside $\Omega$ and the surface $\delta$-function $\delta_\Sigma(\vec{r})$ together with a real function $\lambda(\vec{r})$ defines a contact interaction of the particles with medium, in which the cavity has been formed, at the boundary $\Sigma$. Potential $W(\vec{r}_1, \vec{r}_2)$ describes the interaction between the particles.

From the variational principle with the normalization condition

$$\langle \Psi | \Psi \rangle = \int_\Omega dV_1 dV_2 |\Psi|^2 \quad (3)$$

follows the equation inside the cavity $\Omega$

$$[-\frac{\hbar^2}{2m} \Delta_1 - \frac{\hbar^2}{2m} \Delta_2 + U(\vec{r}_1) + U(\vec{r}_2) + W(\vec{r}_1, \vec{r}_2)] \Psi = E \Psi \quad (4)$$

together with the third type boundary conditions on the cavity surface $\Sigma$

$$\left[ (\vec{n}_1 \vec{\nabla}_1) + \lambda(\vec{r}_1) \right] \Psi(\vec{r}_1, \vec{r}_2) \Big|_{\vec{r}_1 \in \Sigma} = 0$$
$$\left[ (\vec{n}_2 \vec{\nabla}_2) + \lambda(\vec{r}_2) \right] \Psi(\vec{r}_1, \vec{r}_2) \Big|_{\vec{r}_2 \in \Sigma} = 0 \quad (5)$$

where the outward normal to the surface $\Sigma$ is denoted by $\vec{n}$.

If the contact interaction of the particles with cavity environment at $\Sigma$ is absent, $\lambda = 0$ and the conditions (5) transform into the Neumann boundary conditions

$$(\vec{n}_1 \vec{\nabla}_1) \Psi(\vec{r}_1, \vec{r}_2) \Big|_{\vec{r}_1 \in \Sigma} = 0$$
$$(\vec{n}_2 \vec{\nabla}_2) \Psi(\vec{r}_1, \vec{r}_2) \Big|_{\vec{r}_2 \in \Sigma} = 0 \quad (6)$$

The boundary conditions (6) does not necessarily require the presence of particles inside the cavity $\Omega$. Moreover, they could provide the periodic continuation of the wavefunction corresponding to the opposite situation, when the particles are delocalized. For example, in the Wigner-Seitz model of an alkaline metal such delocalization of the particles gives rise to metallic bonding [20]. Therefore, such a "confinement" state of the quantum particles in the cavity with the Neumann boundary conditions is of special interest because in ordered structures such vacuum cavities of the same type could form a spatial lattice. However, in this case the presence of the contact interaction is required because the particles are placed in the cavity formed in a medium and the contact interaction of the particles with environment exists. And if $\lambda \to \infty$, then (5) turn into the Dirichlet boundary conditions

$$\Psi(\vec{r}_1, \vec{r}_2) \Big|_{\vec{r}_1 \in \Sigma} = 0, \ \Psi(\vec{r}_1, \vec{r}_2) \Big|_{\vec{r}_2 \in \Sigma} = 0 \quad (7)$$

and so describe confinement by impenetrable barrier.

## 3. Two-Electron Atoms with $Z \geq 2$ in the Spherically Symmetric Cavity with the Third Type Boundary Conditions

As an example of a two-particle system one can consider a two-electron atom with atomic number $Z$. One can assume that the point nucleus is arranged in the center of a spherically symmetric cavity of radius $R$ and the contact interaction of the electrons with cavity environment is defined by a constant $\lambda = const$. In the Hartree-Fock approximation the form of ansatz for the ground state of the two-electron atom could be written as follows

$$\Psi(\vec{r}_1, \vec{r}_2) = \phi(r_1) \phi(r_2) = \frac{u(r_1) u(r_2)}{r_1 r_2} \quad (8)$$

where $\phi(r)$ corresponds to one-electron wavefunction. It should be noted that this ansatz does not allow obtaining accurate values of the ground state energy, but it is adequate for the qualitative study of the lowest energy level of the two-electron atom with $Z \geq 2$ in a confinement state as a function of cavity spatial parameters.

In what follows, in order to simplify the notation, the relativistic units will be used: $\hbar = c = 1$, wavenumber and energy will be expressed in units of the particle mass $m$, while distances in units of the particle Compton length $1/m$. In these units the energy functional (1) is given by

$$E[u] = \int_0^R dr \left\{ [u'(r)]^2 - \frac{2Z\alpha}{r} u^2(r) \right\} + 2\alpha \int_0^R dr_2 \frac{u^2(r_2)}{r_2} \int_0^{r_2} dr_1 u^2(r_1) + \left( \lambda - \frac{1}{R} \right) u^2(R) \quad (9)$$

with the normalization condition

$$\int_0^R dr \, u^2(r) = 1 \quad (10)$$

The boundary conditions (5) on the cavity surface $\Sigma$ transform into

$$R u'(R) + (\lambda R - 1) u(R) = 0 \quad (11)$$

Now the direct variational method can be applied to obtain the energy of the ground state. For this purpose, the interval $[0, R]$ is divided into $N$ parts, and the energy functional (9) turns into a discrete one. With the boundary condition (11) one obtains instead of the energy functional (9) an algebraic function of the $N-1$ variables. Using the conjugate gradient method under the condition (10) one can minimize the algebraic function to determine the ground state energy of the two-electron atom with sufficient accuracy for a qualitative description of its behavior depending on the parameters $R$ and $\lambda$ at $Z \geq 2$. For example, for the atom with $Z = 2$ in the cavity with the Dirichlet boundary conditions (7) the difference between the ground state energy obtained by using the method described above, and the energy obtained by a more precise calculation in [9], is less than $2\, eV$.

Now let's investigate the asymptotic behavior of the lowest energy level as a function of the cavity spatial parameter $R$ for fixed $\lambda$ at $R \to 0$. It should be noted that in order to remain within the framework of a purely Schroedinger approximation it is necessary to limit the minimal cavity size by $R \sim 10$ [24]. Therefore, the limit of $R \to 0$ is to be understood either as a purely mathematical operation, or as the substitution of $R \sim 10$. According to [22-24] a wavefunction of the one-electron atom in the cavity with the contact potential at the boundary inside the cavity corresponds to the solution of the Schroedinger equation for bound states of an electron in the potential field of a fixed point nucleus. Therefore, at $R \to 0$ the function in the ansatz (5) could be considered as a constant. In this case, with the normalization condition (10) the lowest energy level behaves asymptotically at $R \to 0$ as follows

$$\hat{E}_{2el}(Z, R) = \frac{3(\lambda - Z\alpha)}{R} + \frac{6}{5}\frac{\alpha}{R} \qquad (12)$$

It is easily seen from (12) that the lowest energy level behaves substantially different depending on the value of $\lambda$ at $R \to 0$. Most specific behavior of the lowest energy level at $R \to 0$ will be in case $\lambda = \lambda^s_{2el}(Z)$, where $\lambda^s_{2el}(Z) = (Z - 2/5)\alpha$ – in this case the ground state energy converges to a finite value. At $\lambda > \lambda^s_{2el}(Z)$, the ground state energy increases when $R$ decreases, and vice versa at $\lambda < \lambda^s_{2el}(Z)$.

The asymptotic behavior of the lowest level at $R \to \infty$ corresponds to the results obtained in [22-24]. So, depending on the values of $\lambda$ two types of the asymptotic behavior of the lowest level are possible. The first type takes place when the lowest level exponentially fast approaches the ground state energy of the two-electron atom $E_{2el}(Z)$ (such levels will be denoted as $E_{2el}(Z, R)$). And the second type takes place when the lowest level approaches $\tilde{E}_{2el}(Z) \leq E_{2el}(Z)$ (from this point such levels will be denoted as $\tilde{E}_{2el}(Z)$). The last type is possible only when the contact potential attracts electrons. In this case, at $R \to \infty$ electrons are localized near the cavity boundary, and the Coulomb electron-electron and electron-nucleus interactions become negligible in comparison with the boundary attraction. Therefore, for $\lambda = const$ the asymptotic value $\tilde{E}_{2el}(Z)$ corresponds to the doubled ground state energy of an electron in the spherical cavity at $R \to \infty$

$$\tilde{E}_{2el}(Z) = -\lambda^2 \qquad (13)$$

where $\lambda \leq \lambda^{crit}_{2el}(Z) < 0$ and $\lambda^{crit}_{2el}(Z) = -\sqrt{|E_{2el}(Z)|}$. The asymptotic behavior of those levels should be power.

Indeed, as shown in Fig. 1 for $\lambda = [4(3-x)/5]\alpha$ with $x = 0, 1.., 5$, which are bigger than $\lambda^{crit}_{2el}(Z)$, and for $Z = 2$ exponential levels $E_{2el}(Z, R)$ approach $E_{2el}(Z)$ with increasing $R$, where $E_{2el}(Z) = -(Z - 5/16)^2 \alpha^2$ in the approximation (8). Also it is seen that the greater the deviation of $\lambda$ from the value $\lambda^s_{2el}(Z)$, the slower exponential levels $E_{2el}(Z, R)$ converge to $E_{2el}(Z)$ with increasing $R$. On the other hand, for $x = 2,..,5$ the binding energy of the two-electron atom increases when $R$ decreases and at $R$ of the order of $a_B / 2$ ($a_B \simeq 137$ - Bohr radius) is several times bigger than the binding energy of the free atom. For $x = 1$ which corresponds to $\lambda = \lambda^s_{2el}(Z)$ is shown that in most specific case the lowest level $E_{2el}(Z, R)$ converges to $E_{2el}(Z)$ at $R \to 0$. It should be noted that in Fig. 1 the minimum size of the cavity is limited from bellow $R \geq 10$ to remain within the framework of the purely Schroedinger approximation. The same limitation is presented in all subsequent figures.

Next, one can explore the stability of the two-electron atom with atomic number $Z \geq 2$ in the cavity. For further discussion some information about the asymptotic behavior of the one-electron atom lowest level is required. From [23], for example, it is known that at $R \to 0$ the one-electron atom lowest level behaves as

$$\hat{E}_{1el}(Z, R) = \frac{3}{2}\frac{(\lambda - Z\alpha)}{R} \qquad (14)$$

It is easily seen from (14) that the most specific behavior of the one-electron atom lowest level, when it converges at $R \to 0$ to a finite value, will be in case $\lambda = \lambda^s_{1el}(Z)$, where $\lambda^s_{1el}(Z) = -Z\alpha$. In fact, $E_{1el}(Z, R) = E_{1el}(Z)$ for such value of $\lambda$ and any $R$, where $E_{1el}(Z) = -(Z\alpha)^2 / 2$ is the ground state energy of the free one-electron atom with atomic number $Z$. At $\lambda > \lambda^s_{1el}(Z)$ the ground state energy increases when $R$ decreases, and vice versa at $\lambda < \lambda^s_{1el}(Z)$. At $R \to \infty$ the exponential levels $E_{1el}(Z, R)$ approach $E_{1el}(Z)$, and the power levels $E_{1el}(Z, R)$ approach

$$\tilde{E}_{1el}(Z) = -\lambda^2 / 2 \qquad (15)$$

where $\lambda \leq \lambda^{crit}_{1el}(Z) < 0$ and $\lambda^{crit}_{1el}(Z) = -\sqrt{2|E_{1el}(Z)|}$.

It follows from (12) and (14) that for $\lambda = (Z-4/5)\alpha$

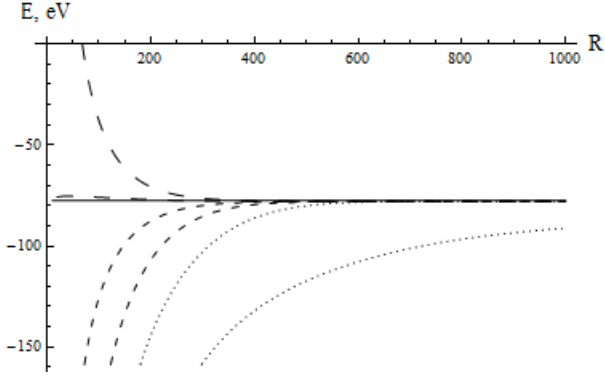

**Figure 1.** *The lowest level of the two-electron atom with $Z=2$ as a function of cavity radius $R$ for $\lambda = [4(3-x)/5]\alpha$ with $x=0, 1$ (long-dashed line), 2, 3 (short-dashed line), 4, 5 (dotted line); $E_{2el}(Z)$ (solid line)*

the asymptotes $E_{1el}(Z,R)$ and $E_{2el}(Z,R)$ coincide, i.e. when the binding energy of the one-electron atom should be smaller than the binding energy of two-electron atom with the same $Z$ at $R \to \infty$, and vice versa when $\lambda > (Z-4/5)\alpha$. Thus, for $\lambda > (Z-4/5)\alpha$ always exists such a cavity, although small in size, in which the two-electron atom decays into the one-electron atom and an electron. This fact is explicitly shown in Fig. 2a, where the behavior of the one- and two-electron atoms for $\lambda = (3/2)\lambda_{2el}^{s}(Z)$ is displayed. For $\lambda < (Z-4/5)\alpha$ the situation is opposite – the ionization energy of two-electron atoms grows with decreasing $R$.

On the other hand, the exponential level $E_{2el}(Z,R)$ converges to $E_{2el}(Z)$, and the power level $E_{2el}(Z,R)$ converges to $E_{2el}(Z) \le E_{2el}(Z)$ at $R \to \infty$. In particular, for $\lambda > (Z-4/5)\alpha$ with increasing $R$ the curves $E_{2el}(Z,R)$ and $E_{1el}(Z,R)$ intersect, and the ionization energy of the two-electron atom converges to the value $E_{1el}(Z) - E_{2el}(Z)$ at $R \to \infty$. For $\lambda < (Z-4/5)\alpha$ the two-electron atom remains stable for any $R$. And for $\lambda_{2el}^{crit}(Z) < \lambda < (Z-4/5)\alpha$ the lowest levels of the one- and two-electron atoms are exponential, and the ionization energy of the two-electron atom with increasing $R$ converges to $E_{1el}(Z) - E_{2el}(Z)$. For $\lambda_{1el}^{crit}(Z) < \lambda \le \lambda_{2el}^{crit}(Z)$ only the lowest level of the two-electron atom is already power, and for $\lambda \le \lambda_{1el}^{crit}(Z)$ not only the level of the two-electron atom is already power but also the level of the one-electron one. In this case the ionization energy of the two-electron atom with increasing $R$ slowly approaches the values $E_{1el}(Z) - \lambda^2$ and $\lambda^2/2$, respectively, as shown in Fig. 2b for $\lambda_{2el}^{crit}(Z)$ and $\lambda_{1el}^{crit}(Z)$ at $Z=2$.

Thus, for $\lambda > (Z-4/5)\alpha$ it is possible that the two-electron atom in the cavity decays into the one-electron atom and an electron. At the same time, for $\lambda < (Z-4/5)\alpha$ the ionization and binding energies of the atoms in the cavity could be several times bigger than the same energies of the free ones, and hence the such cavities are effective containers for the two-electron atoms. Moreover, for $\lambda < \lambda_{1el}^{crit}(Z)$ the binding and ionization

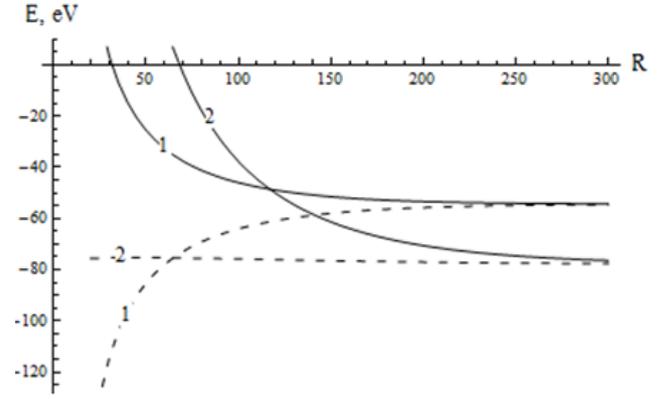

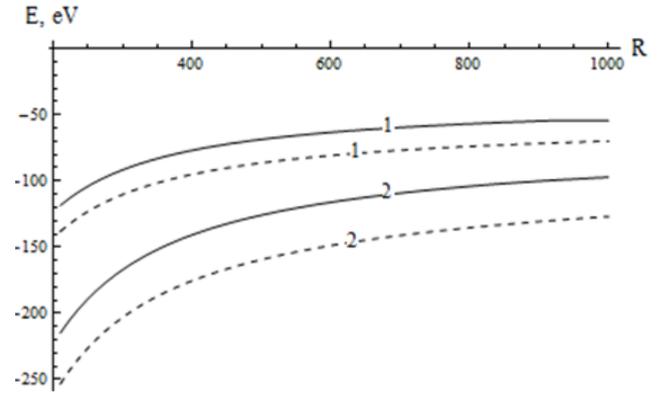

**Figure 2.** *The lowest levels of the one-electron atom (1) and of the two-electron atom (2) with $Z=2$ as a function of cavity radius $R$ for $\lambda = (3/2)\lambda_{2el}^{s}(Z)$ (solid line) and $\lambda = \lambda_{2el}^{s}(Z)$ (dashed line) (a); for $\lambda = \lambda_{2el}^{crit}(Z)$ (solid line) and $\lambda = \lambda_{1el}^{crit}(Z)$ (dashed line) (b)*

energies of the two-electron atom will be bigger than the same energies of the free atom even at $R \to \infty$.

## 4. Two-Electron Atoms with $Z \ge 2$ in the Cavity with an Outer Shell

Up to this point it was assumed that atomic electrons interact with cavity environment only at its boundary $\Sigma$ by means of $\delta$-like potential. In a more realistic approach one should consider instead of a $\delta$-like interaction an outer potential shell of nonvanishing width $d$, into which the electrons penetrate and interact there with cavity environment. In the simplest case, the outer shell potential could be approximated by a constant $V_0$, i.e. one should assume that the interaction between electrons and nucleus and the Coulomb repulsion between electrons are completely screened in the shell. Such a potential is related to atomic physics and is used, for example, for simulating the $C_n$ cage in the electronic structure studies of endohedral fullerenes [15, 25]. The constant $V_0$ should depend on the shell width $d$ so that in the limit $d \to 0$ such a potential shell should transform into contact

interaction at the boundary $\Sigma$

$$V_0 d \to \frac{\hbar^2}{2m}\lambda \,, \quad d \to 0 \qquad (16)$$

Notice that the limits $d \to 0$ and $V_0 \to \infty$ don't commute, because when $d$ is finite and one obtains the confinement of the atom by an impermeable barrier instead of the contact interaction at the boundary $\Sigma$.

In the spherically symmetric case with a constant potential in the shell instead of boundary conditions (5) one obtains an equation of Schroedinger type

$$\left[-\frac{\hbar^2}{2m}\Delta_1 - \frac{\hbar^2}{2m}\Delta_2 + 2V_0\right]\Psi = E\Psi, \qquad (17)$$
$$R \le r \le X = R + d$$

with the Neumann conditions on the outward shell boundary

$$\left.\frac{\partial}{\partial r_1}\Psi(\vec{r}_1,\vec{r}_2)\right|_{r_1=X} = 0 \,, \quad \left.\frac{\partial}{\partial r_2}\Psi(\vec{r}_1,\vec{r}_2)\right|_{r_2=X} = 0 \qquad (18)$$

In such a model, the interaction of electrons with cavity environment is defined by the equation (17), so $\lambda$ is absent in (18). The boundary conditions (18) does not necessarily require the presence of particles inside the cavity $\Omega$. Therefore, if real microcavity might be approximated by a spherically symmetric cavity with an outer shell and the Neumann boundary conditions, the ground state of charged particles in a cubic lattice, formed by cavities of the same type, could be described as in the Wigner-Seitz model [21]. The well-known example of such lattices is given by octahedral and tetrahedral interstitial sites in certain metals and alloys [26-28]. The cavity together with the outer shell forms a kind of the Wigner-Seitz cell, while the boundary conditions (18) provide a periodic continuation of the wavefunctions between neighboring cells.

In view of the above the energy functional (9) transforms into the form

$$E[u] = \int_0^X dr \left[u'(r)\right]^2 - 2Z\alpha \int_0^R dr \frac{u^2(r)}{r} +$$
$$2\alpha \int_0^R dr_2 \frac{u^2(r_2)}{r_2}\int_0^{r_2} dr_1 u^2(r_1) + 2V_0 \int_R^X dr\, u^2(r) - \frac{u^2(X)}{X} \qquad (19)$$

and the normalization condition (10) takes the form

$$\int_0^X dr\, u^2(r) = 1 \qquad (20)$$

The Neumann boundary conditions (18) turn into

$$X u'(X) = u(X) \qquad (21)$$

Now let's investigate the asymptotic behavior of the lowest energy level as a function of the cavity and shell spatial parameters $R$ and $d$ at $R \to 0$ and $d \to 0$. In this case, the limits $R \to 0$ and $d \to 0$ are also to be understood as either purely mathematical operation, or as the substitution of $R \sim 10$ and $d \sim 10$. As in the previous section, at $R \to 0$ the function $\phi(r)$ in the ansatz (5) inside the cavity could be considered as a constant. The function $\phi(r)$ in the outer shell at $R \to 0$ corresponds to the problem of an electron inside the potential barrier of finite height $V_0$ and width $d$ with the Neumann condition on the outer shell boundary. Therefore, in the shell $\phi(r)$ at $R \to 0$ and $d \to 0$ could also be considered as a constant. This means that the lowest energy level behaves asymptotically with the normalization condition (20) at $R \to 0$ as follows

$$\hat{E}_{2el}(Z,R) = -3Z\alpha \frac{R^2}{X^3} + 2V_0\left(1 - \frac{R^3}{X^3}\right) + \frac{6}{5}\alpha\frac{R^5}{X^6} \qquad (22)$$

and the limit of (22) at $R \to 0$ is the value $2V_0$.

There are two types of asymptotic behavior of the lowest level at $R \to \infty$ in dependence on outer shell parameters in analogy with the previous section. The limit of the exponential lowest levels $E_{2el}(Z,R)$ is $E_{2el}(Z)$ and the limit of the power lowest levels $E_{2el}(Z,R)$ is $E_{2el}(Z) \le E_{2el}(Z)$, which corresponds to the doubled ground state energy of an electron in the attractive potential $V_0 < 0$

$$\tilde{V}(r) = \begin{cases} 0, & 0 < r < R \\ V_0, & R \le r < X \end{cases} \qquad (23)$$

at $R \to \infty$. Thus, according to [23] one can obtain the limit $E_{2el}(Z)$ by solving the equation

$$\sqrt{2|V_0| + \tilde{E}_{2el}(Z)} \tan\left(\sqrt{2|V_0| + \tilde{E}_{2el}(Z)}\, d\right) = \sqrt{-\tilde{E}_{2el}(Z)} \qquad (24)$$

Now one can investigate the asymptotic behavior of the lowest energy level as a function of the cavity spatial parameter $R$ using the numerical minimization results of the energy functional (19) under the conditions (20) and (21) for fixed $V_0$ and $d$. Parameters of the outer shell are selected to correspond to the scales of the actual conditions of microcavities, in which such a confinement state is possible (from this point the energy potentials are expressed in $eV$). The values of $|V_0|$ could vary from $\sim 1\, eV$ in superfluid helium [29] and $\sim 5 \div 10\, eV$ for interstices in metal lattices [27, 28] up to dozens $eV$ in quantum chemistry [4, 14-18]. For example, as one of the characteristic values of $V_0$ could be chosen the positive average value $V_0 = 20\, eV$. In this case the ground state energy coincides with the exponential lowest level. Also another case of interest arises when in the outer shell acts the "critical" potential $V_0 = V_{2el}^{crit}(Z)$ and the lowest level turns out to be the power one $E_{2el}(Z,R)$. The values of such a potential can be determined by solving (24) with the substitution $E_{2el}(Z) = \tilde{E}_{2el}(Z)$. Since the calculations are performed within the approximation (8) then the ground state energy of the free two-electron atom $\tilde{E}_{2el}(Z) = -(Z-5/16)^2 \alpha^2$. The numerical values of $V_{2el}^{crit}(Z)$ for $d = x \cdot a_B$ with $x = 1/4, 1/2, 1, 2$ and $Z = 2$ are presented in Tab. 1.

**Table 1.** The values of $V_{2el}^{crit}(Z=2)$ for $d = x a_B$

| $x = d/a_B$ | 1/4 | 1/2 | 1 | 2 |
|---|---|---|---|---|
| $V_{2el}^{crit}(Z), eV$ | -119.74 | -74.59 | -53.07 | -43.86 |

For $Z = 2$ and the outer shell width $d = x \cdot a_B$ with $x = 1/4, 1/2, 1, 2$ for $V_0 = 20\ eV$ there are minima in curves $E_{2el}(Z, R)$ and the minimum depth increases while $d$ decreases as shown in Fig. 3a. For example, for $Z = 2$ and $d = a_B / 4$ the minimum is clearly pronounced, but it is already pronounced rather weakly at $d = a_B / 2$. And for $d = a_B, 2a_B$ the maximal binding energy of the two-electron atom will be attained at large cavity radii $R \gg a_B$. As a consequence while filling a spatial lattice, formed by such a microcavities, with two-electron atoms, the bulk compression or extension is possible as it occurs upon hydrogenation of some metals [26-28]. Furthermore, as it is shown in Fig. 3a the limits of the curves $E_{2el}(Z, R)$ and of the asymptotes (22) coincide and are equal to $2V_0$ at $R \rightarrow 0$. Fig. 3b shows that for $V_0 = V_{2el}^{crit}(Z)$ the difference between $E_{2el}(Z) = \tilde{E}_{2el}(Z)$ and the power lowest levels decrease sufficiently slower than that of the exponential ones, which already arrive at $\tilde{E}_{2el}(Z)$ for $R$ of the order of several $a_B$ as shown in figure 3a. And at $R$ of the order of $a_B / 2$ the binding energy of the two-electron atom is several times bigger than the energy of the free one for $V_0 = V_{2el}^{crit}(Z)$ and $d = a_B / 4, a_B / 2$.

Next, one can explore the stability of the two-electron atom with atomic number $Z \geq 2$ in the cavity with the spherical outer shell of nonvanishing width. For further discussion some information about the asymptotic behavior of the one-electron atom lowest level is required. From [23], for example, it is known that at $R \rightarrow 0$ the limit of the one-electron atom lowest level is $V_0$ for any $d$. At $R \rightarrow \infty$ the exponential lowest levels $E_{1el}(Z, R)$ converge to $E_{1el}(Z)$ and the power ones $E_{1el}(Z, R)$ arrive to $\tilde{E}_{1el}(Z) \leq E_{1el}(Z)$, which for attractive potential $V_0 < 0$ can be defined from the equation

$$\sqrt{2|V_0| + 2\tilde{E}_{1el}(Z)} \tan\left(\sqrt{2|V_0| + 2\tilde{E}_{1el}(Z)}d\right) = \\ = \sqrt{-2\tilde{E}_{1el}(Z)} \qquad (25)$$

The corresponding values of the "critical" potential $V_{1el}^{crit}(Z)$ for the one-electron atoms, while the lowest level turns out to be the power one, and for $Z = 1$ and $d = x \cdot a_B$ with $x = 1/4, 1/2, 1, 2$ are presented in Tab. 2.

**Table 2.** The values of $V_{1el}^{crit}(Z=2)$ for $d = x a_B$

| $x = d/a_B$ | 1/4 | 1/2 | 1 | 2 |
|---|---|---|---|---|
| $V_{1el}^{crit}(Z), eV$ | -148.19 | -95.13 | -70.24 | -59.89 |

Considering the above, when $V_0 > 0$ for any $d$ at $R \rightarrow 0$ the lowest energy level of the two-electron atom will lie above the lowest level of the one-electron atom and vice versa for $V_0 < 0$.

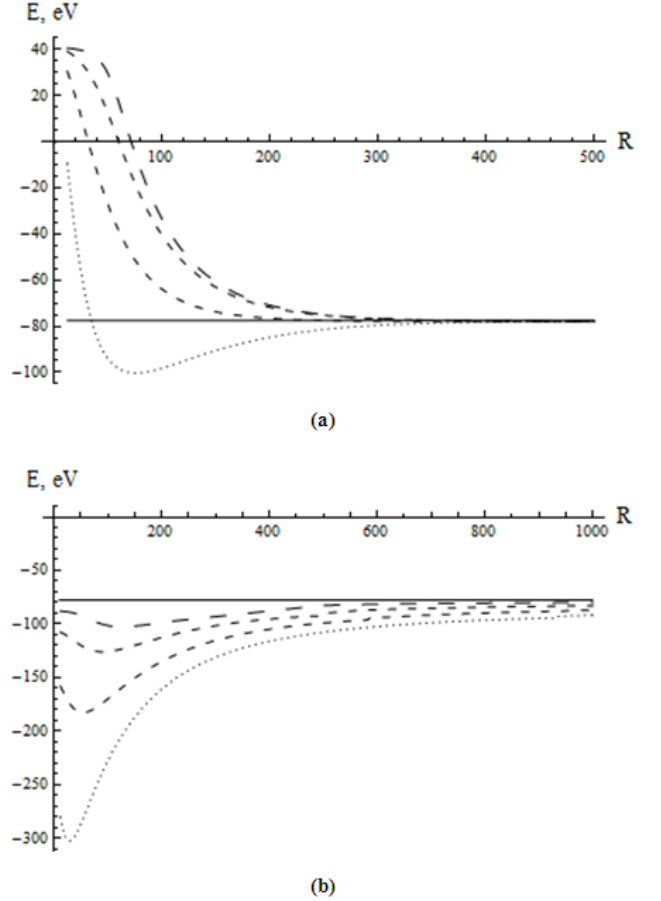

**Figure 3.** The lowest levels of the two-electron atom with $Z = 2$ as a function of cavity radius $R$ at $V_0 = 20\ eV$ (a) and $V_0 = V_{2el}^{crit}(Z)$ (b) for $d = x \cdot a_B$ with $x = 1/4$ (dotted line), $1/2, 1$ (short-dashed line), $2$ (long-dashed line); $\tilde{E}_{2el}(Z)$ (solid line)

On the other hand, taking into account (24), (25), at $R \rightarrow \infty$ for any $V_0$ and $d$ the lowest energy level of the two-electron atom will lie below the lowest level of the one-electron atom.

Numerical calculations show that for any fixed $V_0 > 0$ curves $E_{2el}(Z, R)$ and $E_{1el}(Z, R)$ intersect and this intersection point moves toward the large values of $R$, while $d$ increases, as shown in Fig. 4. Furthermore, in dependence on the outer shell parameters $V_0 > 0$ and $d$ it is possible that the two-electron atom in the cavity decays into the one-electron atom and an electron (see Fig. 4b) or vice versa the ionization and binding energies of such an two-electron atom could be substantially bigger than the same energies of the free one (see Fig. 4a). In particular, it is shown in Fig. 4a that the difference between the maximum binding and ionization energies of the two-electron atom in the cavity and the same energies of the free atom is of the order of several dozens $eV$ for $V_0 = 20\,eV$ and $d = a_B/4$.

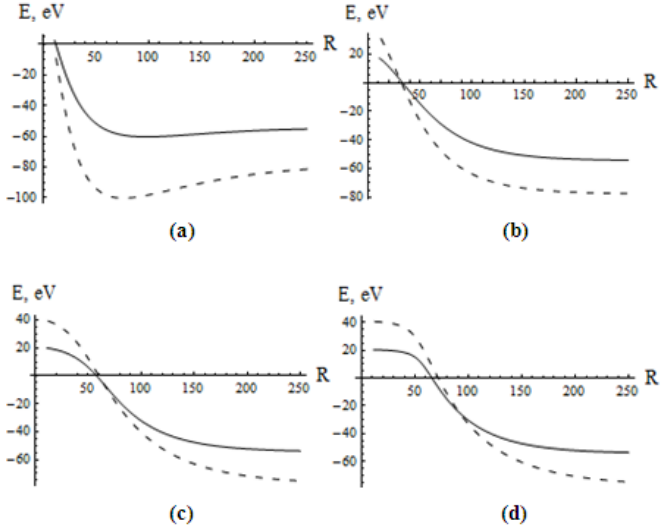

*Figure 4. The lowest levels of the one-electron atom (solid line) and of the two-electron atom (dashed line) with $Z=2$ as a function of cavity radius $R$ at $V_0 = 20\,eV$ for $d = a_B/4$ (a), $d = a_B/2$ (b), $d = a_B$ (c) and $d = 2a_B$ (d)*

When $V_0 < 0$ the lowest level of the two-electron atom will always lie below the level of the corresponding one-electron atom. And for $V_{2el}^{crit}(Z) < V_0 < 0$ the lowest levels of the one- and two-electron atoms are exponential, and the ionization energy of the two-electron atom converges rapidly with increasing $R$ to $E_{1el}(Z) - E_{2el}(Z)$.

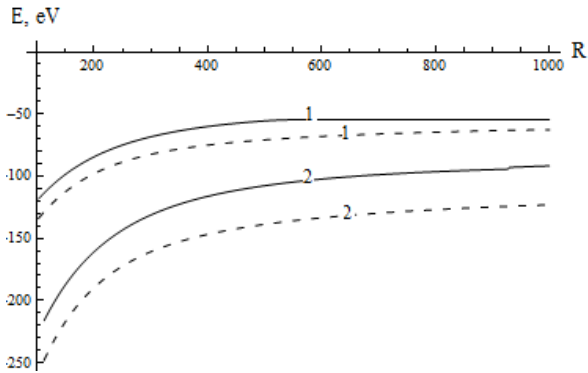

*Figure 5. The lowest levels of the one-electron atom (1) and of the two-electron atom (2) with $Z=2$ as a function of cavity radius $R$ for $d = a_B/4$ at $V_0 = V_{2el}^{crit}(Z)$ (solid line) and $V_0 = V_{1el}^{crit}(Z)$ (dashed line)*

For $V_{1el}^{crit}(Z) < V_0 < V_{2el}^{crit}(Z)$ the lowest level of the two-electron atom is already power and the one of the one-electron atom is still exponential. In this case, the ionization energy of the two-electron atom converges slowly with increasing $R$ to $E_{1el}(Z) - E_{2el}(Z)$. And for $V_0 \leq V_{1el}^{crit}(Z)$ both levels become power and the ionization energy slowly converges to $E_{1el}(Z)$. Both of these cases are displayed in Fig. 5 for $V_0 = V_{1el}^{crit}(Z)$ and $V_0 = V_{2el}^{crit}(Z)$ at $d = a_B/4$.

Thus, for $V_0 > 0$ there are always the outer shell parameters $V_0$, $R$ and $d$ that the two-electron atom in a cavity decays into the corresponding one-electron atom and an electron. On the other hand, there is such a set of the parameters that the binding and ionization energies of the two-electron atom in the cavity could be several times bigger than the same energies of the free one, and hence such cavities are effective traps for the two-electron atoms. And for $V_0 < V_{2el}^{crit}(Z)$ the binding and ionization energies of the two-electron atom are bigger than the same energies of the free atom even at $R \to \infty$.

## 5. Conclusion

To conclude it should be mentioned that the ansatz (8) is adequate for a qualitative studying the binding and ionization energies behavior of a two-electron atom in dependence on the cavity parameters for $Z \geq 2$. For $Z = 1$ in general the considered method is not sufficiently accurate even for qualitative studies. For example, in this approximation the ground state energy of the free two-electron atom with $Z = 1$ lies above the ground state energy of the corresponding one-electron atom ($E_{2el}(Z=1) \approx 0.95 \times E_{1el}(Z=1)$), so that one cannot even conclude that the free two-electron atom can be in the bound state. More precise calculations [30, 31] show that $E_{2el}(Z=1) \approx 1.05 \times E_{1el}(Z=1)$. However, for the cavity parameters, for which the binding energy of the two-electron atom with $Z = 1$ is several times bigger than the binding energy of the free one, the approximation (8) is quite adequate for the qualitative studying the behavior of the binding and ionization energies of the two-electron atom with $Z = 1$ in the cavity. In this case the binding energy of the two-electron atom with $Z = 1$ as a function of the cavity parameters behaves as in the case $Z \geq 2$.

In the present work the influence of the boundary conditions of "not going out" on the ground state properties of the two-electron atom in a cavity was studied. The state of "not going out" from the cavity was provided in two ways: in the first case by means of the $\delta$-shaped potential on the cavity boundary and in the second more realistic case by means of the potential outer shell of nonvanishing width with the conditions (18) imposed on the outward surface of the shell. It was shown that in both cases in dependence on the cavity parameters the two-electron atom in the cavity could decay into the corresponding one-electron atom and an electron or could be in the ground state, in which the binding and ionization energies are several times bigger than the same energies of the free atom. The more realistic case with the outer potential shell is of specific interest because the boundary conditions (18) could provide not only the localization of the electrons inside the cavity but also the periodic continuation of the wavefunction between neighboring cavities. As a result, such cavities could form the spatial lattice, as it occurs in the description of the metallic bond in the framework of the Wigner-Seitz model [21]. And in turn, for the certain cavity and shell parameters such a lattice could be an effective container for the two-electron atoms.

## Acknowledgements

The author is grateful to Prof. Konstantin Sveshnikov from Moscow State University for setting up a problem and discussions.